\documentclass[trackchanges]{aastex631}
\usepackage[utf8]{inputenc}
\usepackage{amsmath}
\usepackage{amssymb}
\usepackage{bm} % For bold math symbols
\usepackage{graphicx}
\usepackage[shortlabels]{enumitem}

\submitjournal{The Astronomical Journal}

\shorttitle{Errors in Pseudo-Circular Feeds}
\shortauthors{D. Mitra}

\begin{document}

\title{Systematic Polarization Errors from Parallactic-Angle Dependent
  Leakage in Pseudo-Circular Feeds}

\author[0000-0002-9142-9835]{Dipanjan Mitra}
\affiliation{National Centre for Radio Astrophysics, Tata Institute of Fundamental Research, Pune 411007, India.}

%\maketitle

\begin{abstract}
Wideband radio interferometers increasingly rely on analog quadrature
hybrids to synthesize circular polarization from linear feeds. These
systems are typically calibrated under the assumption that
instrumental polarization leakage can be represented as a static
complex offset, independent of parallactic angle. In this work, we
demonstrate that this assumption breaks down in the presence of
realistic hybrid imperfections.

We show that amplitude and phase errors in the hybrid
$\mathbf{H}(\nu)$ introduce a non-commutative interaction with
parallactic rotation $\mathbf{R}(\chi)$, such that [
  $\mathbf{H}(\nu)\mathbf{R}(\chi) \neq
  \mathbf{R}(\chi)\mathbf{H}(\nu)$] leading to a time-dependent
effective leakage term that rotates in the Stokes $(Q,U)$ plane. This
effect causes systematic distortions in polarization angle and
introduces frequency-dependent biases that can mimic or corrupt
Faraday rotation measurements.

We derive a first-order analytic model for this leakage and
demonstrate that it manifests as a deterministic, geometrically
modulated error proportional to total intensity. To mitigate this
effect, we introduce a Static Offset Pre-correction (SOP) method that
operates in the antenna frame, inverting the hybrid response prior to
parallactic de-rotation. Unlike conventional calibration approaches,
SOP removes the non-commutative error in the Jones domain, preventing
its projection into the sky frame.

Our results show that hybrid-induced leakage is not merely a
calibration artifact but a fundamental systematic error that must be
addressed to achieve high-fidelity wideband polarimetry.
\end{abstract}
\keywords{Magnetic fields --- Methods: analytical --- Methods: data
  analysis --- Techniques: interferometric --- Techniques:
  polarimetric --- Radio Interferometry --- Rotation Measure}

\section{Introduction}

High-fidelity broadband polarimetry is a definitive diagnostic of the
physical state of magnetized plasmas across astrophysical scales. In
the radio regime, the primary observable is the Faraday rotation of
linearly polarized emission, which provides a unique probe of the
magneto-ionic medium
\citep[e.g.,][]{1966ARA&A...4..245G,2020Galax...8...53H}. This
effect---characterized by a rotation of the Electric Vector Position
Angle (EVPA) proportional to $\lambda^2$---encodes the integral of the
magnetic field and electron density along the line of sight. Capturing
this information requires accurate polarimetric measurements across a
wide parameter space: fractional polarizations ($m$) can range from
$<1\%$ to nearly $100\%$, while Rotation Measures (RM) vary from a few
to several thousand rad m$^{-2}$ \citep{Burn1966, Brentjens2005}.

Early developments in the field transitioned from narrow-band,
single-frequency measurements to the modern era of wide-band
spectro-polarimetry \citep[for a detailed history,
  see][]{2012JAHH...15...76W}.  While early observations were often
limited by bandwidth depolarization, current digital backends provide
the frequency resolution necessary for RM Synthesis and Faraday
Tomography. At low frequencies ($<2$~GHz), where Faraday effects are
most pronounced, measurements are further complicated by ionospheric
Faraday rotation, which can introduce time-varying EVPA rotations of
tens of radians \citep{Sault1996b}.

The response of a radio interferometer is commonly described using the
Radio Interferometer Measurement Equation (RIME), in which
instrumental effects are represented as Jones matrices acting on the
incident electric field \citep{Hamaker1996}. Within this framework,
polarization leakage is typically modeled as a small,
direction-independent perturbation that can be treated as a static
complex offset and removed through calibration. A key implicit
assumption in this approach is that instrumental corruptions commute
with geometric transformations, such as parallactic angle rotation,
allowing the separation of sky rotation from instrumental response.

 To simplify calibration in alt-azimuth mounted systems, many
  interferometers operate in a circular polarization basis, in which
  parallactic angle rotation is absorbed into the receptor response
  and does not affect the parallel-hand visibilities (RR and LL) in an
  ideal system. Instead, the effect of parallactic angle appears as a
  phase rotation in the cross-hand visibilities (RL and LR), with
  opposite signs in $V_{\rm RL}$ and $V_{\rm LR}$, typically of the
  form $e^{\pm 2 i \chi}$. In standard calibration pipelines, this
  geometric phase term is commonly tracked or removed during an
  initial calibration step (often via polarization or
  parallactic-angle correction), allowing subsequent calibration to
  proceed in a frame where the sky polarization is effectively
  de-rotated and the residual calibrated visibilities are
  approximately invariant to parallactic angle.  Because broadband
circular feeds are technically difficult to realize, this basis is
often synthesized from native linear receptors using analog quadrature
hybrids \citep{Thompson2017}. This ``pseudo-circular'' architecture
has been widely adopted across multiple facilities, particularly in
low-frequency and wide-band receivers.

In such systems, however, the instrumental response is defined in the
antenna frame, while parallactic rotation acts as a coordinate
transformation between the sky and the receptors. As a result, the
hybrid response $\mathbf{H}(\nu)$ and the parallactic rotation operator
$\mathbf{R}(\chi)$ do not, in general, commute:
\begin{equation}
\mathbf{H}(\nu)\mathbf{R}(\chi) \neq \mathbf{R}(\chi)\mathbf{H}(\nu).
\end{equation}
This inequality is the formal origin of the effective leakage term
derived in this work.

Realistic quadrature hybrids exhibit frequency-dependent amplitude
imbalances $\epsilon(\nu)$ and phase tracking errors $\delta(\nu)$,
arising from structural dispersion and impedance mismatches within the
receiver chain. While these defects are static in the antenna frame,
their non-commutative interaction with parallactic rotation transforms
them into a time-dependent leakage term in the sky frame. Specifically,
the resulting leakage vector rotates in the Stokes $(Q,U)$ plane at
twice the parallactic angle, coupling total intensity into the
cross-hand visibilities.

This effect has important observational consequences. The induced
leakage introduces systematic distortions in the recovered EVPA and
generates frequency-dependent structure that can mimic or bias Faraday
rotation signals. Because standard calibration procedures assume a
time-invariant leakage model, these errors cannot be fully removed
using conventional techniques based on unpolarized calibrators. The
result is a residual systematic that propagates into Rotation Measure
estimates, particularly in wide-band observations.

In this paper, we formalize this non-commutative interaction and derive
a first-order analytic model for hybrid-induced leakage in
pseudo-circular systems. We show that the resulting error manifests as
a deterministic, geometrically modulated term proportional to total
intensity. To mitigate this effect, we introduce a Static Offset
Pre-correction (SOP) method that operates in the antenna frame,
inverting the hybrid response prior to parallactic de-rotation. This
approach removes the non-commutative error at the voltage level,
restoring the validity of standard calibration assumptions and
enabling high-fidelity wide-band polarimetric recovery.

\section{Visibility Matrix for Linear Feeds}

The interferometric response of a dual-linear array is described using
the Radio Interferometer Measurement Equation (RIME), in which all
propagation and instrumental effects are represented by Jones matrices
acting on the sky coherency matrix.

The incoming radiation undergoes:
1) astrophysical Faraday rotation in the source or intervening medium;
2) ionospheric Faraday rotation along each antenna line of sight;
3) geometric projection into the antenna frame through the parallactic angle;
4) instrumental corruption through complex gains and polarization leakage.

\subsection{Full Jones-chain representation}

Throughout this section, uppercase indices $(A,B)$ denote the two
antennas forming an interferometric baseline, while the generic index
$p$ refers to an arbitrary antenna in the array.
The visibility measured on baseline $AB$ is written as
\begin{equation}
\mathbf{V}_{AB}
=
\mathbf{J}_A
\mathbf{R}(\chi_A)
\mathbf{F}_{\rm ion}^{A}
\mathbf{F}_{\rm FR}
\mathbf{C}_0
\mathbf{F}_{\rm FR}^{\dagger}
\left(\mathbf{F}_{\rm ion}^{B}\right)^{\dagger}
\mathbf{R}^{\top}(\chi_B)
\mathbf{J}_B^{\dagger},
\label{eq:full_rime}
\end{equation}
where:
\begin{itemize}
\item $\mathbf{C}_0$ is the intrinsic sky coherency matrix,
\item $\mathbf{F}_{\rm FR}=\mathbf{R}(\phi_{\rm FR})$ represents astrophysical Faraday rotation,
\item $\mathbf{F}_{\rm ion}^{p}=\mathbf{R}(\phi_{\rm ion}^{p})$ is the ionospheric Faraday rotation for antenna $p$,
\item $\mathbf{R}(\chi_p)$ is the parallactic angle rotation,
\item $\mathbf{J}_p$ is the instrumental Jones matrix.
\end{itemize}

The rotation matrix in the linear polarization basis is
\begin{equation}
\mathbf{R}(\theta)
=
\begin{pmatrix}
\cos\theta & \sin\theta \\
-\sin\theta & \cos\theta
\end{pmatrix}.
\end{equation}

The intrinsic sky coherency matrix is
\begin{equation}
\mathbf{C}_0
=
\frac{1}{2}
\begin{pmatrix}
I+Q & U+iV \\
U-iV & I-Q
\end{pmatrix}.
	\label{ant_corr1}
\end{equation}

The instrumental response is fully contained in the antenna-based
Jones matrices,
\begin{equation}
\mathbf{J}_p
=
\begin{pmatrix}
G_{pX} & D_{pX} \\
D_{pY} & G_{pY}
\end{pmatrix},
\end{equation}
where $G_{pX},G_{pY}$ are the complex gains of the two linear receptors
and $D_{pX},D_{pY}$ describe first-order polarization leakage.

\subsection{Absorption of astrophysical Faraday rotation}

The astrophysical Faraday rotation $\phi_{\rm FR}$ is common to all
antennas and therefore acts as a global rotation of the incoming
polarization state. It can consequently be absorbed into the intrinsic
sky coherency matrix:
\begin{equation}
\mathbf{C}
=
\mathbf{F}_{\rm FR}
\mathbf{C}_0
\mathbf{F}_{\rm FR}^{\dagger}.
	\label{ant_corr2}
\end{equation}

The astrophysical Faraday rotation is absorbed into the
frequency-dependent sky coherency matrix and therefore does not appear
as a separate operator in the reduced interferometric measurement
equation.

The measurement equation therefore reduces to
\begin{equation}
\mathbf{V}_{AB}
=
\mathbf{J}_A
\mathbf{R}(\chi_A)
\mathbf{F}_{\rm ion}^{A}
\mathbf{C}
\left(\mathbf{F}_{\rm ion}^{B}\right)^{\dagger}
\mathbf{R}^{\top}(\chi_B)
\mathbf{J}_B^{\dagger}.
\label{eq:rime_reduced}
\end{equation}

\subsection{Differential ionospheric Faraday rotation}

In the linear polarization basis, both parallactic-angle rotation and
Faraday rotation are represented by the same form of rotation matrix,
and the total rotation for antenna $p$ may be written as
\begin{equation}
\psi_p
=
\chi_p+\phi_{\rm ion}^{p}.
\end{equation}

We define the differential ionospheric Faraday rotation between the
two antennas as
\begin{equation}
\phi
\equiv
\phi_{\rm ion}^{B}
-
\phi_{\rm ion}^{A}.
\end{equation}

For compact arrays we approximate
\begin{equation}
\chi_A \simeq \chi_B \equiv \chi,
\end{equation}
so that
\begin{equation}
\psi_B
=
\psi_A+\phi.
\end{equation}

Substituting into Eq.~(\ref{eq:rime_reduced}) gives
\begin{equation}
\mathbf{V}_{AB}
=
\mathbf{J}_A
\mathbf{R}(\chi)
\mathbf{C}
\mathbf{R}^{\top}(\chi+\phi)
\mathbf{J}_B^{\dagger}.
	\label{vab_lin}
\end{equation}

\subsection{Response to an unpolarized source}

For an unpolarized source, $Q=U=V=0$,
so that
\begin{equation}
\mathbf{C}
=
\frac{I}{2}
\mathbf{I}_2,
\end{equation}
where $\mathbf{I}_2$ is the $2\times2$ identity matrix.

Since the coherency matrix of an unpolarized source is proportional
to the identity matrix, it commutes with the rotation operators.

Using the orthogonality of rotation matrices,
\begin{equation}
\mathbf{R}(\chi)\mathbf{R}^{\top}(\chi+\phi)
=
\mathbf{R}(-\phi),
\end{equation}
the visibility given by Eq.~(\ref{vab_lin}) becomes
\begin{equation}
\mathbf{V}_{AB}
=
\frac{I}{2}
\mathbf{J}_A
\mathbf{R}(-\phi)
\mathbf{J}_B^{\dagger}.
\label{eq:compact_linear}
\end{equation}

Equation~(\ref{eq:compact_linear}) shows that for an unpolarized
source, differential ionospheric Faraday rotation appears as the only
explicit baseline-dependent propagation term in the interferometric
observable, while astrophysical Faraday rotation remains absorbed into
the sky coherency matrix and the parallactic angle cancels as a common
geometric rotation.

Retaining only first-order leakage terms gives
\begin{align}
V_{XX}
&\approx
\frac{I}{2}
\Big[
G_{AX}G_{BX}^{*}\cos\phi
-
G_{AX}D_{BX}^{*}\sin\phi
+
D_{AX}G_{BX}^{*}\sin\phi
\Big],
\\
V_{XY}
&\approx
\frac{I}{2}
\Big[
-
G_{AX}G_{BY}^{*}\sin\phi
+
G_{AX}D_{BY}^{*}\cos\phi
+
D_{AX}G_{BY}^{*}\cos\phi
\Big],
\\
V_{YX}
&\approx
\frac{I}{2}
\Big[
G_{AY}G_{BX}^{*}\sin\phi
+
G_{AY}D_{BX}^{*}\cos\phi
+
D_{AY}G_{BX}^{*}\cos\phi
\Big],
\\
V_{YY}
&\approx
\frac{I}{2}
\Big[
G_{AY}G_{BY}^{*}\cos\phi
+
G_{AY}D_{BY}^{*}\sin\phi
-
D_{AY}G_{BY}^{*}\sin\phi
\Big].
\end{align}

In the ideal limit of unity gains and zero leakage,
\begin{equation}
V_{XX}+V_{YY}
=
I\cos\phi.
\end{equation}

\subsection{Physical interpretation}

In the unpolarized limit, Eq.~(\ref{eq:compact_linear})
demonstrates that interferometers are sensitive only to differential
propagation effects. The
astrophysical Faraday rotation modifies the intrinsic complex linear
polarization of the sky and remains encoded in the coherency matrix,
while the parallactic angle cancels as a common geometric rotation
between antennas.

The observable effect arises from the differential ionospheric
Faraday rotation $\phi$, which rotates the linear polarization basis
between antennas and mixes signal between co-polar
($V_{XX},V_{YY}$) and cross-polar ($V_{XY},V_{YX}$) correlations.
The cross-hand response scales as $\sin\phi$, while the co-polar
response scales as $\cos\phi$.

To first order, instrumental leakage and differential Faraday rotation
remain quasi-separable: leakage introduces additive mixing of Stokes
$I$ into the cross-hands, whereas differential Faraday rotation
introduces a coherent basis rotation between polarization states.

\section{Visibility Matrix for Native Circular Feeds}

We now consider a baseline between two antennas, $A$ and $B$,
equipped with native circular feeds ($R,L$). As in the linear-feed
case, the incoming radiation undergoes astrophysical Faraday rotation,
ionospheric Faraday rotation, parallactic-angle rotation, and
instrumental corruption.

\subsection{Full Jones-chain representation}

The visibility matrix is written as
\begin{equation}
\mathbf{V}_{AB}
=
\mathbf{J}_A
\mathbf{R}(\chi_A)
\mathbf{F}_{\rm ion}^{A}
\mathbf{F}_{\rm FR}
\mathbf{C}_0
\mathbf{F}_{\rm FR}^{\dagger}
\left(\mathbf{F}_{\rm ion}^{B}\right)^{\dagger}
\mathbf{R}^{\dagger}(\chi_B)
\mathbf{J}_B^{\dagger},
\end{equation}
where $\mathbf{C}_0$ is the intrinsic sky coherency matrix,
$\mathbf{F}_{\rm FR}$ represents astrophysical Faraday rotation,
$\mathbf{F}_{\rm ion}^{p}$ is the ionospheric Faraday rotation for
antenna $p$, and $\mathbf{J}_p$ is the instrumental Jones matrix.

In the circular polarization basis, rotation operators are diagonal:
\begin{equation}
\mathbf{R}(\theta)
=
\begin{pmatrix}
e^{-i\theta} & 0 \\
0 & e^{i\theta}
\end{pmatrix}.
\end{equation}

The intrinsic sky coherency matrix in the circular basis is
\begin{equation}
\mathbf{C}_0
=
\frac{1}{2}
\begin{pmatrix}
I+V & Q+iU \\
Q-iU & I-V
\end{pmatrix}.
\end{equation}

\subsection{Absorption of astrophysical Faraday rotation}

As in the linear-feed formulation, the astrophysical Faraday rotation
is common to all antennas and can therefore be absorbed into the
frequency-dependent sky coherency matrix,
\begin{equation}
\mathbf{C}
=
\mathbf{F}_{\rm FR}
\mathbf{C}_0
\mathbf{F}_{\rm FR}^{\dagger}.
\end{equation}

The observable visibility therefore depends only on antenna-dependent
rotations.

\subsection{Differential ionospheric Faraday rotation}

Defining the total rotation angle for antenna $p$ as
\begin{equation}
\psi_p
=
\chi_p+\phi_{\rm ion}^{p},
\end{equation}
the visibility becomes
\begin{equation}
\mathbf{V}_{AB}
=
\mathbf{J}_A
\mathbf{R}(\psi_A)
\mathbf{C}
\mathbf{R}^{\dagger}(\psi_B)
\mathbf{J}_B^{\dagger}.
\end{equation}

We define the differential ionospheric Faraday rotation
\begin{equation}
\phi
\equiv
\phi_{\rm ion}^{B}
-
\phi_{\rm ion}^{A},
\end{equation}
and for compact arrays assume
\begin{equation}
\chi_A \simeq \chi_B \equiv \chi.
\end{equation}

Thus,
\begin{equation}
\psi_B
=
\psi_A+\phi,
\end{equation}
\subsection{Response to an unpolarized source}
For an unpolarized source,
\begin{equation}
Q=U=V=0,
\end{equation}
so that
\begin{equation}
\mathbf{C}
=
\frac{I}{2}\mathbf{I}_2.
\end{equation}

The rotation matrix then becomes
\begin{equation}
\mathbf{R}(\psi_A)\mathbf{R}^{\dagger}(\psi_B)
=
\begin{pmatrix}
e^{i\phi} & 0 \\
0 & e^{-i\phi}
\end{pmatrix}.
\end{equation}

The measurement equation for an unpolarized source therefore reduces to
\begin{equation}
\mathbf{V}_{AB}
=
\mathbf{J}_A
\begin{pmatrix}
e^{i\phi} & 0 \\
0 & e^{-i\phi}
\end{pmatrix}
\mathbf{C}
\mathbf{J}_B^{\dagger}.
\label{eq:circular_compact}
\end{equation}

Equation~(\ref{eq:circular_compact}) shows that in the native circular
basis, differential Faraday rotation appears purely as opposite phases
in the two circular polarization channels.

Assuming ideal gains ($G=1$) and zero leakage ($D=0$), the visibilities become
\begin{equation}
V_{RR}
=
\frac{I}{2}e^{i\phi},
\qquad
V_{LL}
=
\frac{I}{2}e^{-i\phi}.
\end{equation}

Thus, for an unpolarized source, the parallactic angle cancels
exactly, while differential Faraday rotation enters only as conjugate
phase terms in the two circular channels.

Unlike the linear-feed case, there is no redistribution of power
between co-polar and cross-polar correlations for an unpolarized 
source. The total intensity
remains entirely in the co-polar visibilities, while Faraday rotation
introduces only phase modulation.
After calibration of the antenna-based phases, Stokes $I$ is recovered
directly from the invariant visibility amplitudes. This diagonal
structure enables a near-separation between propagation-induced
rotation and instrumental leakage. 

The key distinction between the native linear and native circular
bases is therefore that, in the ideal circular basis, geometric and
propagation-induced rotations remain diagonal in the visibility
matrix and do not mix power between polarization channels.
The pseudo-circular systems discussed
in the following section depart from this ideal behavior because the
linear-to-circular conversion is performed by a non-ideal analog
hybrid operating in the antenna frame.

\section{Pseudo-Circular Feeds: Instrument Model and Calibration Implications}

Pseudo-circular feeds are formed by combining orthogonal linear
receptors $(X,Y)$ through an analog quadrature hybrid. In the ideal
case, the hybrid introduces a $90^\circ$ phase offset between the two
linear signal paths, producing orthogonal circular outputs. In
practice, imperfections in both the native linear feeds and the hybrid
introduce residual instrumental polarization that couples to geometric
rotation and modifies the observed visibility response.

\subsection{Measurement Equation and Operator Ordering}

The signal path is naturally described using the RIME formalism as an
ordered sequence of Jones operators acting on the incoming electric
field. The sky coherency matrix is first rotated into the antenna
reference frame, after which the instrumental response of the linear
feeds acts on the voltages. The quadrature hybrid then transforms the
signal from the linear basis into the pseudo-circular basis prior to
correlation.

For antenna $p$, the full response is written as
\begin{equation}
	\mathbf{V}_{AB}^{meas}
=
\mathbf{H}_A
\mathbf{J}_A
\mathbf{R}(\chi_A)
\mathbf{C}
\mathbf{R}^{\top}(\chi_B)
\mathbf{J}_B^\dagger
\mathbf{H}_B^\dagger,
\end{equation}
where:
\begin{itemize}
\item $\mathbf{R}(\chi)$ is the parallactic-angle rotation operator,
\item $\mathbf{J}$ describes the native linear-feed response,
\item $\mathbf{H}$ is the quadrature-hybrid operator,
\item $\mathbf{C}$ is the sky coherency matrix.
\end{itemize}
Since the pseudo-circular system is formed from native linear receptors,
the parallactic-angle rotation operators retain the linear-basis form,
for which $\mathbf{R}^{-1}=\mathbf{R}^{\top}$.

As discussed in the previous section, astrophysical Faraday rotation is
fully encoded in the complex sky polarization $(Q+iU)$ and therefore
does not appear as an explicit operator in the interferometric
measurement equation. Differential ionospheric Faraday rotation is
neglected here in order to isolate the instrumental effects associated
with pseudo-circular feed formation.
Accordingly, only parallactic-angle rotation is retained explicitly in
the following pseudo-circular measurement equations.

The linear-basis rotation matrix is
\begin{equation}
\mathbf{R}(\chi)=
\begin{pmatrix}
\cos\chi & \sin\chi \\
-\sin\chi & \cos\chi
\end{pmatrix},
\end{equation}
and the sky coherency matrix in linear-basis is
\begin{equation}
\mathbf{C}
=
\frac12
\begin{pmatrix}
I+Q & U+iV \\
U-iV & I-Q
\end{pmatrix}.
\label{cmatrix_lin}
\end{equation}

The native linear-feed response is parameterized as
\begin{equation}
\mathbf{J}
=
\mathbf{G}+\mathbf{D},
\end{equation}
with
\begin{equation}
\mathbf{G}=
\begin{pmatrix}
G_X & 0 \\
0 & G_Y
\end{pmatrix},
\qquad
\mathbf{D}=
\begin{pmatrix}
0 & D_X \\
D_Y & 0
\end{pmatrix},
\end{equation}
where $\mathbf{G}$ contains the complex gain and bandpass response and
$\mathbf{D}$ describes first-order polarization leakage between the
orthogonal linear receptors.

\subsection{Antenna-Frame Coherency Matrix}

Before hybrid conversion, the sky polarization is projected into the
antenna frame through parallactic rotation:
\begin{equation}
\mathbf{X}
=
\mathbf{R}(\chi)\,
\mathbf{C}\,
\mathbf{R}^{\top}(\chi).
\end{equation}

Substituting the explicit matrix forms gives
\begin{equation}
\mathbf{X}
=
\frac12
\begin{pmatrix}
\cos\chi & \sin\chi \\
-\sin\chi & \cos\chi
\end{pmatrix}
\begin{pmatrix}
I+Q & U+iV \\
U-iV & I-Q
\end{pmatrix}
\begin{pmatrix}
\cos\chi & -\sin\chi \\
\sin\chi & \cos\chi
\end{pmatrix}.
\end{equation}

Evaluating the matrix products yields
\begin{align}
X_{11}
&=
\frac12
\left(
I + Q\cos2\chi + U\sin2\chi
\right), \\
X_{22}
&=
\frac12
\left(
I - Q\cos2\chi - U\sin2\chi
\right), \\
X_{12}
&=
\frac12
\left(
U\cos2\chi - Q\sin2\chi + iV
\right).
\end{align}

These expressions describe the correlations presented to the native
linear receptors prior to hybrid conversion. Total intensity remains
invariant under rotation, while the linear polarization vector
$(Q,U)$ rotates at twice the parallactic angle.

\subsection{General Parameterization of the Hybrid Jones Matrix}

The imperfect quadrature hybrid is represented by a frequency-dependent
Jones matrix
\begin{equation}
\mathbf{H}(\nu) =
\begin{pmatrix}
H_{XX}(\nu) & H_{XY}(\nu) \\
H_{YX}(\nu) & H_{YY}(\nu)
\end{pmatrix},
\end{equation}
acting on the antenna-frame linear voltages prior to correlation.

In general, the hybrid response may be factorized into three physically distinct
contributions:
\begin{equation}
\mathbf{H}
=
\mathbf{G}_H\,
\mathbf{H}_0\,
(\mathbf{I}+\Delta \mathbf{H}),
\end{equation}
where:
\begin{itemize}
	\item $\mathbf{H}_0$ is the ideal quadrature hybrid,
\begin{equation}
\mathbf{H}_0
=
\frac{1}{\sqrt{2}}
\begin{pmatrix}
1 & i \\
1 & -i
\end{pmatrix},
\label{eq:H0}
\end{equation}
which performs the ideal linear-to-circular basis transformation,
mapping the native linear receptor voltages $(X,Y)$ into the
pseudo-circular basis $(R,L)$.

	\item $\mathbf{G}_H = \mathrm{diag}(g_X^H, g_Y^H)$ contains diagonal
	complex gains and bandpass terms associated with the two hybrid arms.

	\item $\Delta \mathbf{H}$ contains non-unitary departures from ideal
	quadrature behavior.
\end{itemize}

The factorization above separates hybrid imperfections into:
(i) ordinary diagonal complex gains,
(ii) the ideal unitary basis transformation,
and (iii) residual non-unitary perturbations responsible for
polarization mixing.

The diagonal gain matrix $\mathbf{G}_H$ is observationally degenerate
with standard antenna-based gain and bandpass calibration and is
therefore absorbed into conventional interferometric calibration
solutions. Likewise, fixed differential phase slopes between the
pseudo-circular channels are removed through standard RL/LR delay
calibration. Physically, these delays arise from unequal electrical
path lengths between the hybrid outputs, LNAs, and correlator signal
chains. In practice they are measured using either a correlated noise
source or an astronomical calibrator and removed prior to polarization
calibration.

Consequently, after standard calibration, the dominant residual
instrumental effects are the differential amplitude and phase
imbalances contained in $\Delta \mathbf{H}$, which form the focus of
the SOP formalism developed below.

Explicitly, the imperfect hybrid response may be represented
to first order by
\begin{equation}
\mathbf{H} =
\begin{pmatrix}
g_X (1+\epsilon_X)e^{i\delta_X} & \kappa_{XY} \\
\kappa_{YX} & g_Y (1+\epsilon_Y)e^{i(\pi/2+\delta_Y)}
\end{pmatrix},
\end{equation}
where:
\begin{itemize}
	\item $g_X$ and $g_Y$ are complex path gains,
	\item $\epsilon_X,\epsilon_Y$ are fractional amplitude errors,
	\item $\delta_X,\delta_Y$ are small phase perturbations,
	\item $\kappa_{XY},\kappa_{YX}$ describe internal cross-coupling
	between the hybrid arms.
\end{itemize}

The parameterization above provides a phenomenological first-order
description of amplitude imbalance, phase error, and internal
cross-coupling within the hybrid signal paths.
Higher-order non-linearities and
frequency-dependent couplings beyond first order are neglected,
consistent with the small-error regime assumed throughout this work.

Assuming all imperfections remain small,
\begin{equation}
|\epsilon_i|,\ |\delta_i|,\ |\kappa_{ij}| \ll 1,
\end{equation}
the exponential terms may be expanded as
\begin{equation}
e^{i\delta} \approx 1 + i\delta,
\end{equation}
yielding
\begin{align}
H_{XX} &\approx g_X (1 + \epsilon_X + i\delta_X), \\
H_{YY} &\approx g_Y\, i (1 + \epsilon_Y + i\delta_Y).
\end{align}

Factoring out the common scalar response
\begin{equation}
g_X(1+\epsilon_X+i\delta_X),
\end{equation}
the remaining observable structure depends only on differential and
non-unitary terms between the two hybrid arms. The common scalar term
is fully degenerate with standard antenna-based complex gain,
bandpass, and RL/LR delay calibration, and is therefore absorbed into
the conventional calibration chain prior to polarization calibration.

The effective normalized hybrid response may therefore be written as
\begin{equation}
\tilde{\mathbf{H}} =
\begin{pmatrix}
1 & d_{XY} \\
d_{YX} & i(1+\epsilon+i\delta)
\end{pmatrix},
\end{equation}
where
\begin{align}
\epsilon &\equiv \epsilon_Y-\epsilon_X, \\
\delta &\equiv \delta_Y-\delta_X,
\end{align}
and $d_{XY}, d_{YX}$ denote normalized internal cross-coupling terms.

For many practical quadrature hybrids, the dominant residual
non-ideality after standard calibration is the imperfect $90^\circ$
phase relation between the two arms. In this regime, internal
cross-coupling remains sub-dominant,
\begin{equation}
|d_{XY}|,\ |d_{YX}| \ll |\epsilon|,\ |\delta|,
\end{equation}
and the hybrid response reduces to a single effective complex
coefficient describing the perturbed quadrature arm:
\begin{equation}
H_Y \approx i(1+\epsilon+i\delta).
\end{equation}

Expanding to first order,
\begin{equation}
H_Y \approx i + i\epsilon - \delta.
\end{equation}

The apparent asymmetry between the two hybrid arms is purely a
normalization convention: the $X$-arm is chosen as the reference, and
all physically relevant departures are expressed as differential errors
in the orthogonal arm.

\subsection{First-Order Measurement Equation}

After gain normalization, the measurement equation in the
pseudo-circular system becomes

\begin{align}
	\mathbf{V}_{AB}^{\rm meas}
\approx\;&
\mathbf{H}_A
\Big[
(\mathbf{I}+\mathbf{D}_A)\,
\mathbf{R}(\chi_A)\mathbf{C}\mathbf{R}^{\top}(\chi_B)\,
(\mathbf{I}+\mathbf{D}_B^\dagger)
\Big]
\mathbf{H}_B^\dagger
\nonumber\\
&+ \mathcal{O}(\mathbf{D}^2).
\end{align}

The feed-leakage matrices $\mathbf{D}_A$ and $\mathbf{D}_B$
act in the native linear receptor basis and therefore retain the
standard parallactic-angle covariance structure.

The quadrature hybrid acts after parallactic rotation and performs a
basis transformation from the native linear receptor frame into the
pseudo-circular frame. Since the hybrid operator generally does not
commute with parallactic rotation,
\[
\mathbf{H}\mathbf{R}(\chi)\neq \mathbf{R}(\chi)\mathbf{H},
\]
instrumental leakage terms acquire additional polarization mixing in
the pseudo-circular basis.

This non-commutativity arises because $\mathbf{H}$ is defined in the
antenna-based linear receptor basis, whereas $\mathbf{R}(\chi)$
represents a rotation in the sky-projected linear polarization basis.
The mismatch of bases implies that $\mathbf{H}$ is not diagonal in the
rotated frame, producing order-dependent polarization mixing.

Unlike conventional feed leakage ($D$-terms), which transforms
covariantly with parallactic angle in the native linear basis, the
hybrid defect originates in the basis-conversion stage itself and
therefore introduces additional non-commuting polarization mixing in
the pseudo-circular frame.

\subsection{Static-Offset Pre-Correction (SOP)}

To isolate instrumental effects associated with the pseudo-circular
transformation, and assuming the hybrid is deterministic and
invertible up to calibration error, we define the static-offset
pre-correction (SOP)
\begin{equation}
	\mathbf{V}_{AB}^{\rm SOP}
=
\mathbf{H}_A^{-1}
	\mathbf{V}_{AB}^{\rm meas}
\mathbf{H}_B^{-\dagger}.
\end{equation}

Operationally, SOP is applied prior to parallactic-angle derotation so
that the hybrid-induced antenna-frame offset is removed before it can
couple into the sky-frame polarization response.

This operation assumes that $\mathbf{H}$ is non-singular and slowly
varying across the calibration interval. In practice, $\mathbf{H}$ is
estimated per antenna using either laboratory characterization or
astronomical self-calibration methods, and the inversion is applied
per solution interval rather than per instantaneous visibility in
order to suppress noise amplification.

For identical deterministic hybrid responses on both antennas and
accurate inversion of the nominal hybrid operator, the leading-order
hybrid contribution cancels. To first order in feed leakage, the
corrected measurement equation reduces to
\begin{align}
	\mathbf{V}_{AB}^{\rm SOP}
\approx\;&
\mathbf{X}_{AB}
+
\mathbf{D}_A\,\mathbf{X}_{AB}
+
\mathbf{X}_{AB}\,\mathbf{D}_B^\dagger
+
\mathcal{O}(\mathbf{D}^2)
+
\mathcal{O}(\Delta\mathbf{H}),
\end{align}
where
\begin{equation}
\mathbf{X}_{AB}
=
\mathbf{R}(\chi_A)\mathbf{C}\mathbf{R}^{\top}(\chi_B),
\end{equation}
and $\Delta\mathbf{H}$ denotes residual hybrid calibration error.

This transformation removes the basis-changing effect of the
quadrature hybrid, mapping the data back into an effective
linear-feed representation in which standard gain, bandpass, and
leakage calibration techniques apply

\subsection{Effective Leakage in Cross-Hand Visibilities from Hybrid Non-Commutativity}

To illustrate the non-commutativity between the parallactic rotation
operator $\mathbf{R}(\chi)$ and the quadrature hybrid response
$\mathbf{H}(\nu)$, we focus on the cross-hand visibility $V_{RL}$,
since it contains the lowest-order coupling between total intensity
and linear polarization.

For an ideal circular system, the intrinsic sky-frame cross-hand
visibility is
\begin{equation}
V_{RL}^{\rm true}
=
\frac12(Q+iU).
\label{eq:vrl_true}
\end{equation}

After parallactic rotation into the antenna frame, the ideal measured
visibility becomes
\begin{equation}
V_{RL}^{\rm meas}
=
\frac12(Q+iU)e^{-2i\chi}.
\label{eq:vrl_ant_true}
\end{equation}

Instrumental defects in the pseudo-circular conversion introduce an
additional contribution proportional to Stokes $I$. The cross-hands
therefore provide the most sensitive probe of hybrid-induced
polarization leakage.

To isolate the effect, we consider a baseline in which antenna $B$ is
an ideal reference element:
\begin{equation}
G_B = 1,
\qquad
D_B = 0,
\qquad
\epsilon_B = 0,
\qquad
\delta_B = 0.
\end{equation}

The corresponding pseudo-circular coefficients are
\begin{equation}
\tilde{\alpha}_B = 1,
\qquad
\tilde{\beta}_B = -i,
\end{equation}
so that
\begin{equation}
\tilde{\alpha}_B^\ast = 1,
\qquad
\tilde{\beta}_B^\ast = i.
\end{equation}

Substituting these into the general visibility expansion derived in
Appendix~\ref{app_c} gives
\begin{equation}
V_{RL}^{\rm meas}
=
\alpha^A X_{11}
+
i\alpha^A X_{12}
+
\beta^A X_{21}
+
i\beta^A X_{22},
\label{eq:vrl_start}
\end{equation}
where $X_{ij}$ are the antenna-frame coherency terms.

Using the coherency matrix of Eq.~\ref{cmatrix_lin}, and retaining
only first-order hybrid imperfections for antenna $A$, while
neglecting feed leakage and gain residuals for clarity, we write
\begin{equation}
\alpha^A \simeq 1,
\qquad
\beta^A \simeq i(1+\epsilon_A)-\delta_A,
\end{equation}
where $\epsilon_A$ and $\delta_A$ represent the hybrid amplitude and
phase imbalance, respectively.

Substituting into Eq.~\ref{eq:vrl_start} yields
\begin{align}
V_{RL}^{\rm meas}
=&\,
X_{11}
+iX_{12}
+
\left[i(1+\epsilon_A)-\delta_A\right]X_{21}
\nonumber\\
&
+
i\left[i(1+\epsilon_A)-\delta_A\right]X_{22}.
\end{align}

Expanding the coherency matrix and retaining only first-order terms in
$\epsilon_A$ and $\delta_A$ gives
\begin{align}
V_{RL}^{\rm meas}
\approx&
\frac12(Q+iU)e^{-2i\chi}
\nonumber\\
&
-\frac12 I(\epsilon_A+i\delta_A)
\nonumber\\
&
-\frac12(\epsilon_A+i\delta_A)
(Q-iU)e^{+2i\chi}
\nonumber\\
&
+\frac{i}{2}V
\left(
2+\epsilon_A-i\delta_A
\right).
\label{eq:vrl_full}
\end{align}

The first term is the desired sky polarization rotated into the
antenna frame. The second term represents leakage of total intensity
into the cross-hand visibility due to imperfections in the quadrature
hybrid.

The third term,
\begin{equation}
-\frac12(\epsilon_A+i\delta_A)
(Q-iU)e^{+2i\chi},
\end{equation}
describes a first-order perturbation of the intrinsic linear
polarization itself. Unlike the leakage proportional to Stokes $I$,
this term does not generate polarization from an unpolarized source,
but instead modifies the existing polarized signal.

For most radio sources,
\begin{equation}
|V|,|Q|,|U| \ll I,
\end{equation}
so the dominant contamination arises from the total-intensity leakage
term,
\begin{equation}
-\frac12 I(\epsilon_A+i\delta_A).
\end{equation}

The polarization-dependent correction therefore enters only at
relative order
$\mathcal{O}
\left(
\frac{\sqrt{Q^2+U^2}}{I}
\right)$,
and is typically sub-dominant for weakly polarized sources.
Retaining only the dominant systematic contribution therefore gives
\begin{equation}
V_{RL}^{\rm meas}
\approx
\frac12(Q+iU)e^{-2i\chi}
-
\frac12 I(\epsilon_A+i\delta_A).
\label{eq:vrl_meas_simple}
\end{equation}

The first term corresponds to the ideal antenna-frame visibility
(Eq.~\ref{eq:vrl_ant_true}), while the second term is a static hybrid
defect defined in the antenna frame.
For clarity, the derivation is performed using a
single-defective-antenna baseline in which antenna
$B$ is assumed ideal. The resulting leakage term
therefore represents the effective first-order
baseline leakage associated with antenna $A$.

To recover the sky-frame polarization, the standard calibration
procedure applies parallactic-angle de-rotation:
\begin{equation}
V_{RL}^{\rm corr}
=
V_{RL}^{\rm meas}e^{+2i\chi}.
\end{equation}

Substituting Eq.~\ref{eq:vrl_meas_simple} gives
\begin{equation}
V_{RL}^{\rm corr}
\approx
\frac12(Q+iU)
-
\frac12 I(\epsilon_A+i\delta_A)e^{+2i\chi}.
\label{eq:deff_basic}
\end{equation}

Equation~\ref{eq:deff_basic} demonstrates the central result of this
work: a static hybrid defect in the antenna frame appears as a
rotating leakage term after transformation back into the sky frame.

We therefore define the effective sky-frame leakage as
\begin{equation}
D_{\rm eff}(\nu,\chi)
=
(\epsilon_A+i\delta_A)e^{+2i\chi}.
\label{eq:deff}
\end{equation}

Using this definition,
Eq.~\ref{eq:deff_basic} becomes
\begin{equation}
V_{RL}^{\rm corr}
=
V_{RL}^{\rm true}
-
\frac12 I D_{\rm eff}(\nu,\chi),
\end{equation}
where $V_{RL}^{\rm true}$ is given by
Eq.~\ref{eq:vrl_true}.

Expanding Eq.~\ref{eq:deff} explicitly gives
\begin{align}
D_{\rm eff}(\nu,\chi)
=&\,
\epsilon_A(\nu)\cos2\chi
-
\delta_A(\nu)\sin2\chi
\nonumber\\
&
+i
\left[
\epsilon_A(\nu)\sin2\chi
+
\delta_A(\nu)\cos2\chi
\right].
\end{align}

The leakage amplitude remains invariant,
\begin{equation}
|D_{\rm eff}|
=
\sqrt{\epsilon_A^2+\delta_A^2},
\end{equation}
while its phase rotates at twice the parallactic-angle rate.

Thus, although the hybrid defect is static in the antenna frame, its
projection into the sky frame is intrinsically time dependent.

\begin{figure}[h!]
\centering
\includegraphics[width=0.4\linewidth]{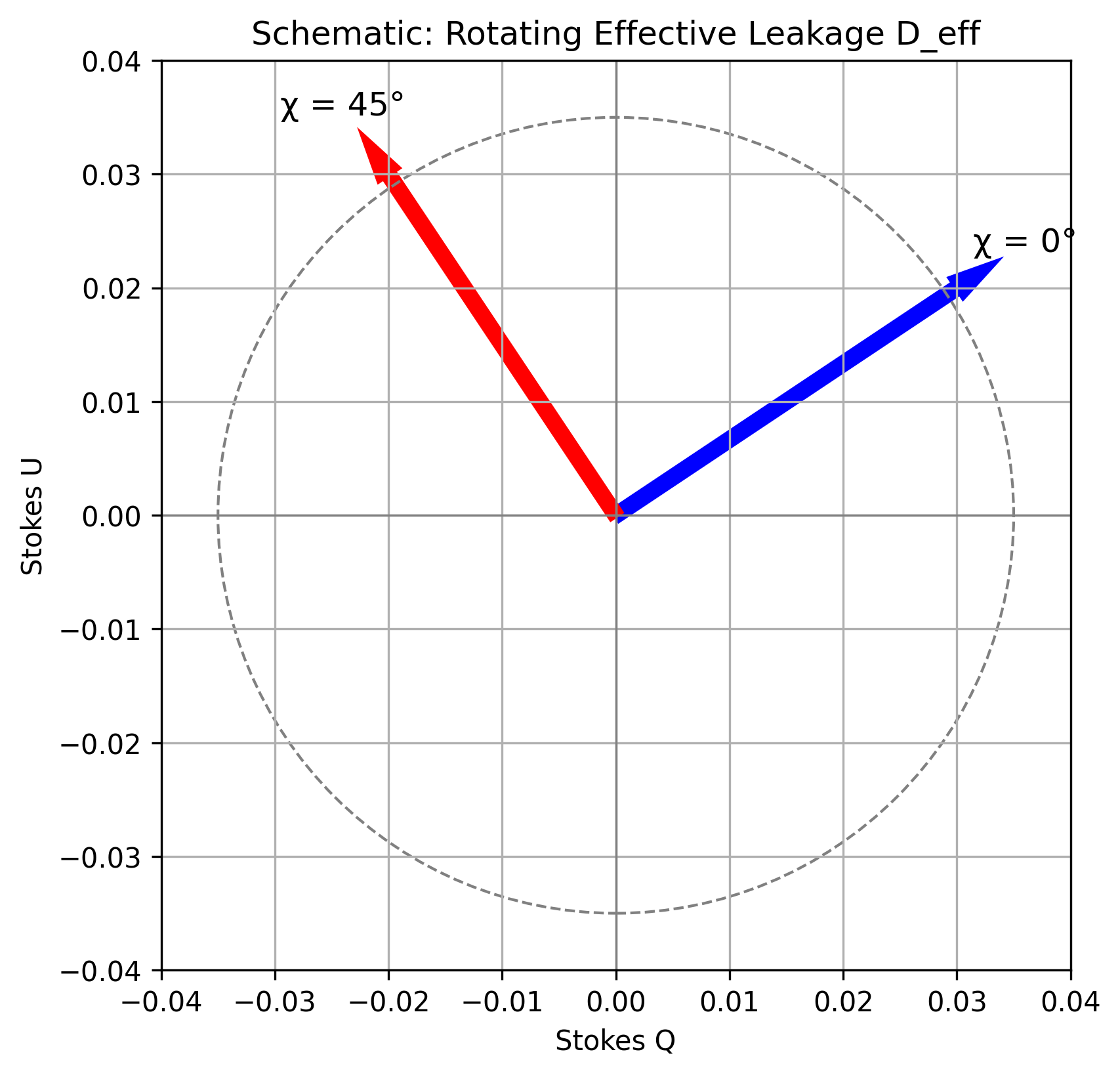}
\caption{
Schematic illustration of the effective leakage vector
$D_{\rm eff}$ in the $(Q,U)$ plane. The leakage amplitude remains
constant while the phase rotates with parallactic angle due to the
non-commutativity between the hybrid response and parallactic
rotation.
}
\label{fig:deff_rotation}
\end{figure}

For example,
\begin{align}
D_{\rm eff}(\nu,0^\circ)
&=
\epsilon(\nu)+i\delta(\nu),
\\
D_{\rm eff}(\nu,45^\circ)
&=
-\delta(\nu)+i\epsilon(\nu),
\end{align}
showing that the real and imaginary components are exchanged (up to a
sign) under parallactic rotation.
This rotation is illustrated in Fig.~\ref{fig:deff_rotation}, which
shows the effective leakage vector rotating in the $(Q,U)$ plane while
its magnitude remains invariant.

This behavior differs fundamentally from that of native circular-feed
systems, where instrumental leakage appears as a static complex offset
in the sky frame. In pseudo-circular architectures, the hybrid
imperfections are introduced prior to parallactic de-rotation,
causing the real and imaginary components of the leakage to mix as the
feed rotates relative to the sky.

Frequency-dependent variations in $\epsilon(\nu)$ and $\delta(\nu)$
therefore generate chromatic rotations of the effective leakage
vector, leading to systematic EVPA distortions and biases in recovered
rotation measures when treated using conventional static-leakage
calibration models.

\subsection{Effective Leakage and RM Bias}

Using the definition derived in
Eq.~\ref{eq:deff},
the de-rotated sky-frame cross-hand visibility may be written as
\begin{equation}
V_{RL}^{\rm corr}
=
V_{RL}^{\rm true}
-
\frac12 I D_{\rm eff}(\nu,\chi),
\label{eq:vrl_corr_rm}
\end{equation}
where
\begin{equation}
V_{RL}^{\rm true}
=
\frac12(Q+iU).
\end{equation}

For a source with fractional polarization $m$ and intrinsic Electric
Vector Position Angle (EVPA) $\zeta_0$, the intrinsic polarization is
\begin{equation}
Q+iU
=
mI e^{2i\zeta_0},
\end{equation}
so that
\begin{equation}
V_{RL}^{\rm true}
=
\frac{mI}{2}e^{2i\zeta_0}.
\end{equation}

Substituting into Eq.~\ref{eq:vrl_corr_rm} gives
\begin{equation}
V_{RL}^{\rm corr}
=
\frac{I}{2}
\left[
m e^{2i\zeta_0}
-
D_{\rm eff}(\nu,\chi)
\right].
\label{eq:vrl_corr_pol}
\end{equation}

The measured EVPA is obtained from the phase of the corrected
cross-hand visibility:
\begin{equation}
\zeta_{\rm meas}
=
\frac12
\arg
\left(
V_{RL}^{\rm corr}
\right).
\end{equation}

Using Eq.~\ref{eq:vrl_corr_pol},
\begin{equation}
\zeta_{\rm meas}
=
\frac12
\arg
\left[
m e^{2i\zeta_0}
-
D_{\rm eff}(\nu,\chi)
\right].
\end{equation}

Assuming the leakage is small compared to the intrinsic polarized
signal,
\begin{equation}
|D_{\rm eff}| \ll m,
\end{equation}
the phase may be linearized by factoring out the intrinsic signal:
\begin{equation}
\zeta_{\rm meas}
=
\zeta_0
+
\frac12
\arg
\left[
1
-
\frac{D_{\rm eff}(\nu,\chi)}{m}
e^{-2i\zeta_0}
\right].
\end{equation}

Applying the small-angle approximation
\begin{equation}
\arg(1+z)\approx \Im(z),
\end{equation}
gives the first-order EVPA error
\begin{equation}
\delta\zeta
\equiv
\zeta_{\rm meas}-\zeta_0
\approx
-
\frac{1}{2m}
\Im
\left[
D_{\rm eff}(\nu,\chi)
e^{-2i\zeta_0}
\right].
\label{eq:evpa_err_general}
\end{equation}

Using
\begin{equation}
D_{\rm eff}(\nu,\chi)
=
(\epsilon+i\delta)e^{2i\chi},
\end{equation}
the EVPA error becomes
\begin{equation}
\delta\zeta
\approx
-
\frac{1}{2m}
\left[
\epsilon(\nu)\sin2(\chi-\zeta_0)
+
\delta(\nu)\cos2(\chi-\zeta_0)
\right].
\label{eq:evpa_err_final}
\end{equation}

Equation~\ref{eq:evpa_err_final} shows that the hybrid-induced leakage
introduces a frequency-dependent rotation of the measured polarization
angle. Since both $\epsilon(\nu)$ and $\delta(\nu)$ are generally
chromatic, the induced EVPA distortion acquires a non-linear
dependence on $\lambda^2$.

For weakly polarized sources, where
\begin{equation}
|D_{\rm eff}| \sim m,
\end{equation}
the leakage-induced phase slope can become comparable to or larger
than the intrinsic Faraday rotation. In this regime, the recovered
Rotation Measure (RM) becomes strongly biased and may even undergo
apparent sign reversals.

\subsection{SOP-Corrected Cross-Hand Visibility and Simulation Framework}

The effective leakage derived above originates from the hybrid defect
being introduced in the antenna frame prior to parallactic
de-rotation. To remove this contribution, we define a
Static-Offset Pre-correction (SOP) scheme in which the hybrid term is
subtracted before applying the geometric rotation correction.

Using the antenna-frame measurement model
\begin{equation}
V_{RL}^{\rm meas}
=
\frac12(Q+iU)e^{-2i\chi}
-
\frac12 I D_{\rm hyb},
\label{eq:sop_meas}
\end{equation}
with
\begin{equation}
D_{\rm hyb}(\nu)
=
\epsilon(\nu)+i\delta(\nu),
\end{equation}
the SOP-corrected visibility is
\begin{equation}
V_{RL}^{\rm SOP}
=
V_{RL}^{\rm meas}
+
\frac12 I D_{\rm hyb}.
\end{equation}

After parallactic-angle de-rotation,
\begin{equation}
V_{RL}^{\rm SOP,corr}
=
V_{RL}^{\rm SOP}e^{2i\chi}
=
\frac12(Q+iU),
\end{equation}
showing that the intrinsic sky polarization is recovered to
first order.

\begin{figure}[ht!]
\centering
\includegraphics[width=\textwidth]{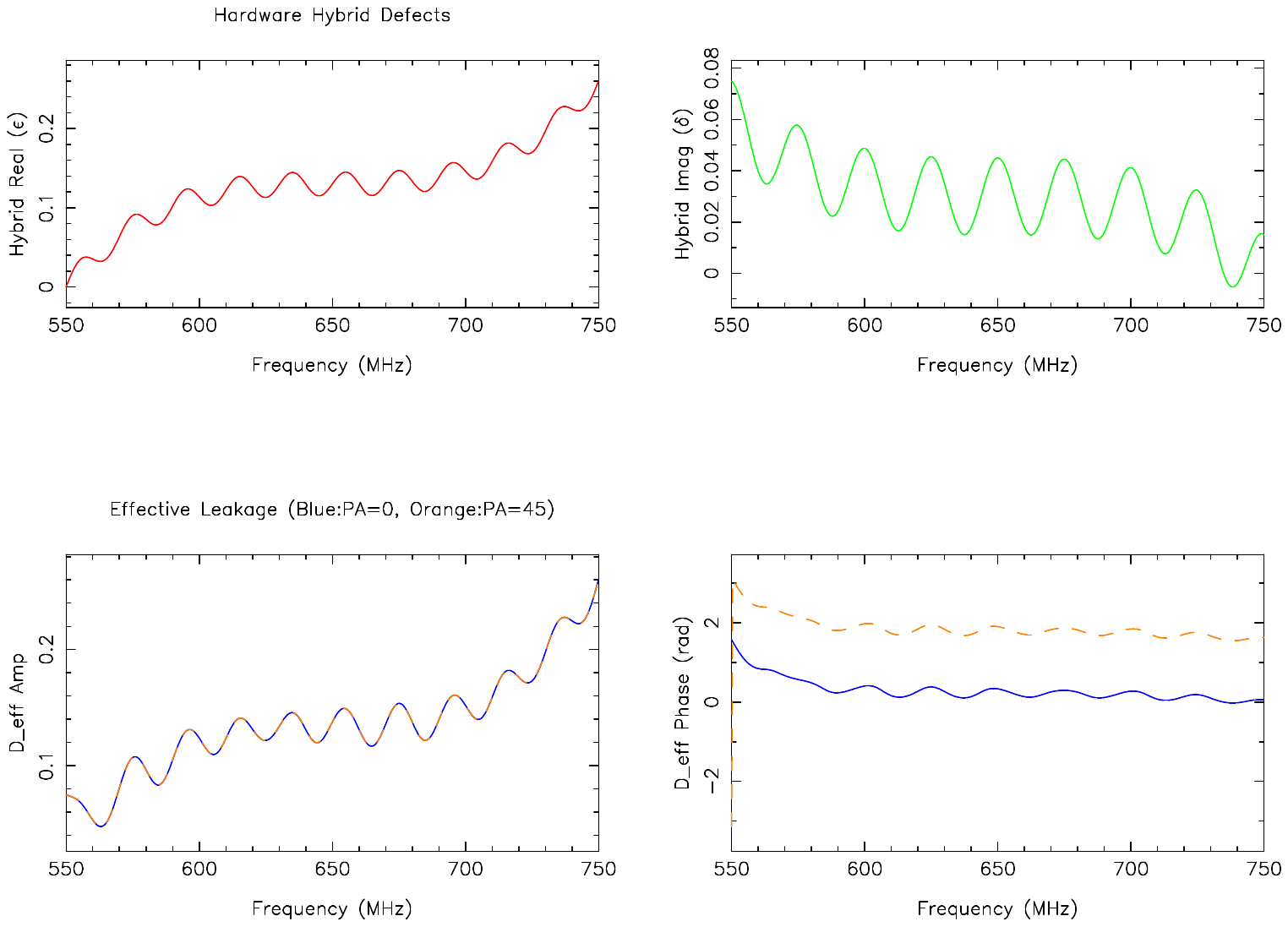}
\caption{Instrumental Hybrid Models and Effective Leakage.  Top panels
  show the simulated frequency-dependent hardware defects of the
  quadrature hybrid: (Left) the amplitude imbalance $\epsilon(\nu)$
  with a cubic baseline and standing-wave oscillations, and (Right)
  the corresponding phase error $\delta(\nu)$.  Bottom panels
  illustrate the effective sky-frame leakage $D_{\rm eff}$ calculated
  at two parallactic angles: $\chi = 0^\circ$ (blue) and $\chi =
  45^\circ$ (orange, dashed).  While the leakage amplitude (Left)
  remains invariant under rotation, the $90^\circ$ phase shift in the
  leakage vector (Right) demonstrates the non-commutative nature of
  the hybrid error.  This rotation of the error vector in the $(Q, U)$
  plane is the primary driver of the systematic EVPA ``wiggles''
  observed in standard calibration pipelines.}
\label{fig:instrument}
\end{figure}

%// Hybrid defect: cubic + wiggles
%void get_hybrid(float f, float f_center, float *eps, float *del) {
%    float off = (f - f_center) / 100.0;
%    float base_eps = 0.13 + 0.13 * pow(off,3.0);
%    float base_del = 0.03 - 0.03 * pow(off,3.0);
%    float wig_eps = 0.015 * sin(2.0*PI*0.05*(f-550.0));
%    float wig_del = 0.015 * cos(2.0*PI*0.04*(f-550.0));
%    *eps = base_eps + wig_eps;
%    *del = base_del + wig_del;
%}

%%%%%%%%%%%%%%%%%%%%%
\begin{figure}[ht!]
\centering

\begin{minipage}[t]{0.49\textwidth}
\centering
{\bf (a)}\\[2mm]
\includegraphics[width=\textwidth,height=10cm]{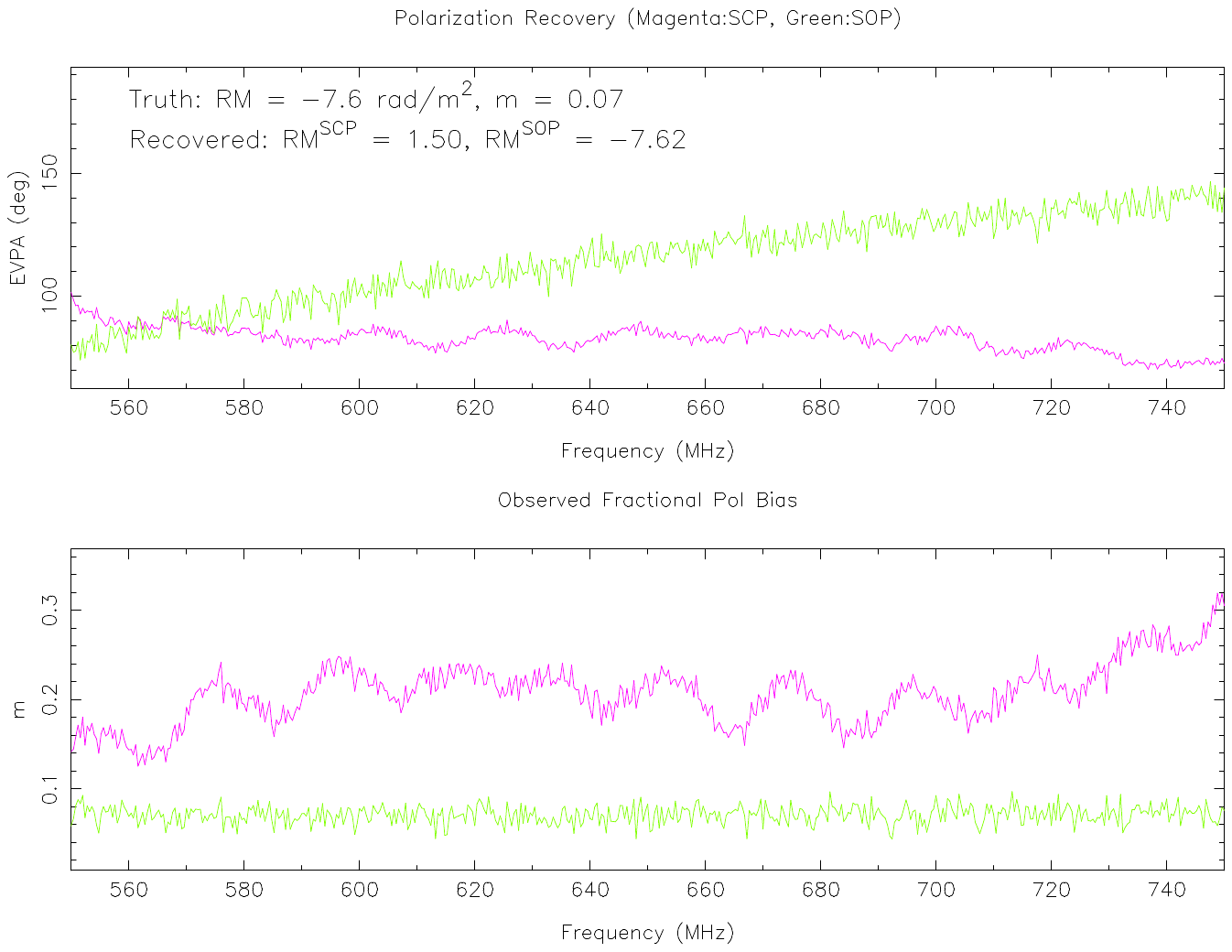}
\end{minipage}
\hfill
\begin{minipage}[t]{0.49\textwidth}
\centering
{\bf (b)}\\[2mm]
\includegraphics[width=\textwidth,height=10cm]{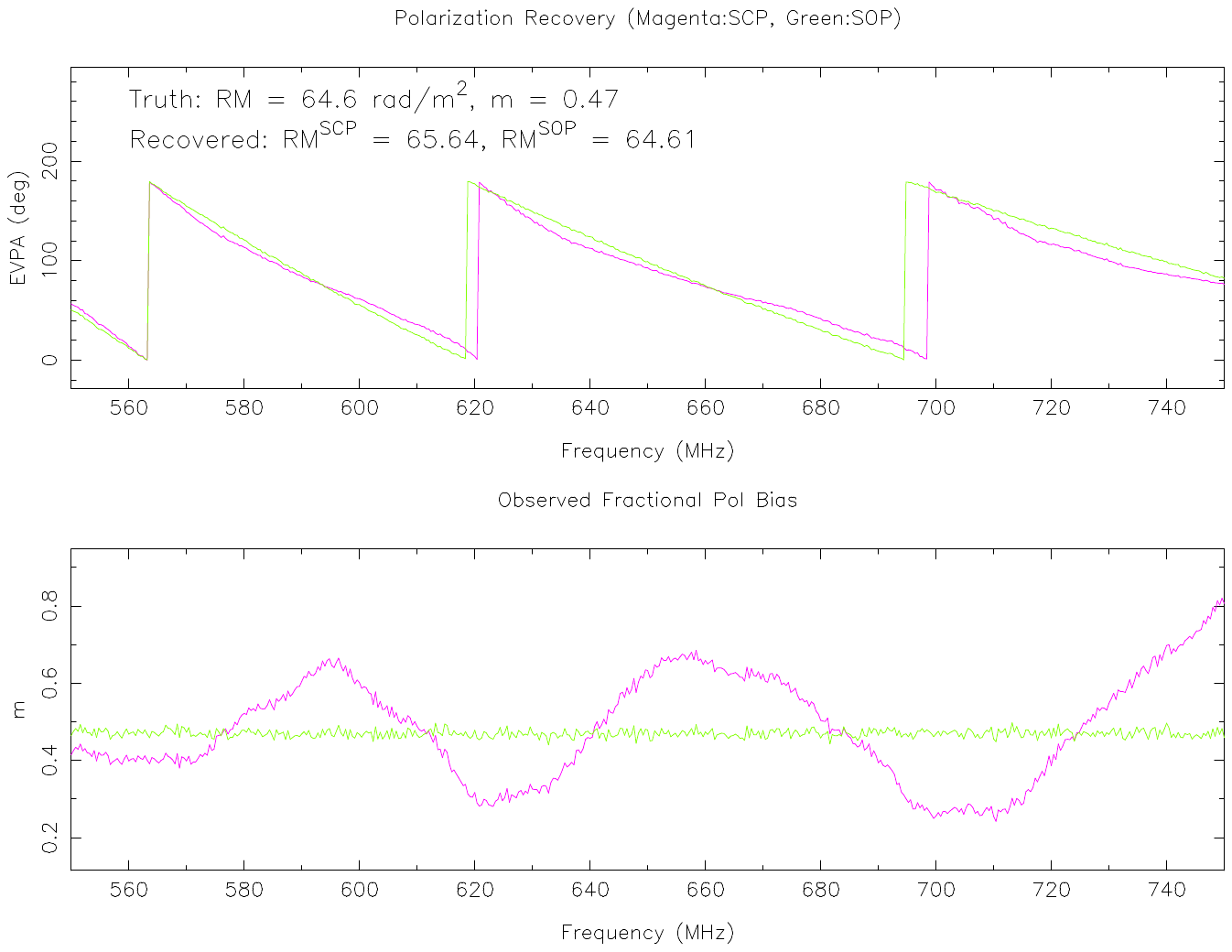}
\end{minipage}

\caption{Comparison of science recovery using SCP and SOP calibration.
(a) Weak-signal regime ($m=0.07$, $\mathrm{RM}=-7.8$~rad~m$^{-2}$).
The hybrid-induced phase slope dominates, producing a significant RM bias
and even a sign reversal.
(b) High-signal regime ($m=0.47$, $\mathrm{RM}=64.6$~rad~m$^{-2}$).
While the recovered RM slope converges more closely to the true value,
the fractional polarization $m$ (bottom panel) exhibits vector-interference oscillations.
In both cases, the SCP solution (magenta) fails to recover the intrinsic behaviour,
whereas the SOP solution (green) accurately reproduces the true signal.}
\label{fig:science}
\end{figure}

%%%%%%%%%%%%%%%%%%%%%%
We simulate broadband observations over 550--750\,MHz using an
intrinsic sky polarization
\begin{equation}
P_{\rm true}
=
Q+iU
=
mI e^{2i(\zeta_0+{\rm RM}\lambda^2)},
\end{equation}
where $\zeta_0=30^\circ$ and $I=1$.

The hybrid imperfections $\epsilon(\nu)$ and $\delta(\nu)$ are modeled
using a cubic baseline combined with standing-wave oscillations, as
shown in Figure~\ref{fig:instrument}; the physical basis for this model 
is discussed in Appendix~\ref{app_d} and is representative of the 
architecture employed in the upgraded GMRT (uGMRT; \citealt{2017CSci..113..707G}).

The measured antenna-frame polarization is then constructed as
\begin{equation}
P_{\rm meas}
=
P_{\rm true}e^{-2i\chi}
-
I D_{\rm hyb}(\nu)
+
n(\sigma),
\end{equation}
where
$n(\sigma)$ represents thermal noise.

After standard parallactic-angle correction, the recovered sky-frame
polarization becomes
\begin{equation}
P_{\rm corr}
=
P_{\rm true}
-
I D_{\rm eff}(\nu,\chi),
\end{equation}
with
\begin{equation}
D_{\rm eff}(\nu,\chi)
=
D_{\rm hyb}(\nu)e^{2i\chi}.
\end{equation}

We compare two calibration strategies:

\begin{enumerate}

\item
Standard calibration proccedure (SCP), in which a static leakage measured at a
reference parallactic angle is subtracted after geometric rotation;

\item
Static-offset pre-correction (SOP), in which the hybrid contribution
is removed directly in the antenna frame before parallactic
de-rotation.

\end{enumerate}

As shown in Figure~\ref{fig:science}a, the SCP method fails in the
weak-polarization regime ($m=0.07$), where the hybrid-induced phase
slope dominates the intrinsic Faraday rotation, producing large RM
biases and even apparent sign reversals.

In the high-signal regime (Figure~\ref{fig:science}b, $m=0.47$), the
recovered EVPA slope approaches the intrinsic Faraday trend because
the source polarization dominates the leakage amplitude. However, the
recovered fractional polarization still exhibits oscillatory structure
due to residual hybrid leakage that is not fully removed by the SCP
approach. These oscillations arise from interference between the
intrinsic polarized signal and the frequency-dependent residual
instrumental term, demonstrating that the recovered polarization
remains systematically contaminated even when the EVPA trend appears
approximately correct.

The SOP method removes the leakage prior to parallactic rotation and
therefore eliminates this residual interference term at first order,
recovering both the intrinsic polarization amplitude and the correct
RM behavior.

The hybrid-induced non-commutativity also affects the recovery of
Stokes $V$. In the high-polarization regime
($m \gg 0.1$), the hybrid phase error $\delta(\nu)$ produces a
first-order coupling between linear and circular polarization,
yielding
\begin{equation}
	{\rm Stokes} ( V_{\rm meas})
\approx
	{\rm Stokes} ( V_{\rm true})
+
mI \sin\delta \sin(2\zeta_{\rm ant}),
\end{equation}
where $\zeta_{\rm ant}=\zeta_0-\chi$ is the orientation of the linear
polarization vector in the antenna frame. Since $\chi$ evolves during
source tracking, this term generates a spurious time-dependent Stokes
$V$ signal.

By contrast, Stokes $I=(V_{RR}+V_{LL})/2$ is affected only at second
order because the dominant hybrid terms enter symmetrically in the
parallel-hand visibilities and cancel to first order. Consequently,
cross-hand and circular-polarization measurements are substantially
more sensitive to hybrid imperfections than total intensity.

For the simulated high-signal case ($m\approx0.5$), a representative
phase error of $\delta=2^\circ$ generates spurious circular
polarization at the level of $\sim2\%$ of Stokes $I$, substantially
larger than the intrinsic circular polarization expected for most
synchrotron sources. The resulting spectral structure follows the
frequency-dependent hybrid response and can mimic astrophysical
circular polarization if uncorrected. Applying SOP suppresses this
term by removing the hybrid phase offset directly in the antenna
frame.

The simulations presented here intentionally assume identical antennas
and deterministic hybrid responses in order to isolate the underlying
non-commutative effect. In practical observations, antenna-based
leakages are often estimated by averaging calibrator observations over
parallactic angle and fitting a static complex D-term. However, in a
pseudo-circular system the effective leakage
$D_{\rm eff}(\nu,\chi)$ is explicitly parallactic-angle dependent, so
the fitted leakage varies with the $\chi$ coverage of the calibrator
observation. Different calibrators therefore yield different effective
leakage solutions, leaving residual frequency-dependent structure even
after standard calibration. The simulations capture the fundamental
origin of this systematic effect.

\subsection{Combined Feed and Hybrid Leakage}

We now generalize the previous derivation by including intrinsic
feed-leakage terms together with the hybrid imperfections. This
separates conventional static instrumental polarization from the
non-commutative leakage introduced by the pseudo-circular hybrid.

Retaining first-order feed leakage terms, the pseudo-circular voltage
coefficients become
\begin{align}
\alpha^A
&=
1 + H_Y D_{YA},
\\
\beta^A
&=
D_{XA}+H_Y,
\end{align}
where
\begin{equation}
H_Y
\approx
i(1+\epsilon_A)-\delta_A.
\end{equation}

Expanding to first order in
$D_{XA}$,
$D_{YA}$,
$\epsilon_A$,
and $\delta_A$ gives
\begin{align}
\alpha^A
&\approx
1+iD_{YA},
\\
\beta^A
&\approx
i
+
D_{XA}
+
i\epsilon_A
-
\delta_A.
\end{align}

Assuming antenna $B$ is ideal, $\tilde{\alpha}_B^\ast = 1$ and $\tilde{\beta}_B^\ast = i$, 
the cross-hand visibility becomes
\begin{equation}
V_{RL}
=
\alpha^A X_{11}
+
i\alpha^A X_{12}
+
\beta^A X_{21}
+
i\beta^A X_{22}.
\end{equation}

Substituting the antenna-frame coherency terms and retaining only
first-order contributions proportional to Stokes $I$ yields the measured visibility:
\begin{equation}
V_{RL}^{\rm meas}
\approx
\frac12(Q+iU)e^{-2i\chi}
+
\frac{i}{2}(D_{XA}+D_{YA})I
-
\frac12(\epsilon_A+i\delta_A)I.
\end{equation}

Defining the conventional static leakage term $D_{\rm static} \equiv i(D_{XA}+D_{YA})$ and the hybrid leakage term $D_{\rm hyb} \equiv \epsilon_A+i\delta_A$ results in
\begin{equation}
V_{RL}^{\rm meas}
\approx
\frac12(Q+iU)e^{-2i\chi}
+
\frac12 I D_{\rm static}
-
\frac12 I D_{\rm hyb}.
\label{eq:vrl_antenna_frame}
\end{equation}

Equation~(\ref{eq:vrl_antenna_frame}) is expressed in the antenna frame. Both
$D_{\rm static}$ and $D_{\rm hyb}$ are antenna-based quantities and
therefore appear as stationary complex offsets in this representation.
In the limit $D_{\rm hyb}=0$, Equation~(\ref{eq:vrl_antenna_frame}) reduces to
the familiar first-order feed-leakage measurement equation used in
conventional polarization calibration.

Standard polarization calibration exploits parallactic-angle
diversity to determine and remove the conventional feed leakage
$D_{\rm static}$. The discussion below therefore concerns the residual
instrumental polarization associated with $D_{\rm hyb}$ after
conventional D-term calibration has been applied.

Equation~(\ref{eq:vrl_antenna_frame}) separates two distinct forms of
instrumental polarization. The conventional feed leakage
$D_{\rm static}$ arises from cross-coupling between the linear
receptors and can therefore be removed through standard polarization
calibration. In contrast, the hybrid leakage $D_{\rm hyb}$ originates
from imperfections in the basis-conversion network.

Because the hybrid response is introduced downstream of the
linear receptors, it is not generally represented by the conventional
feed-leakage model. Residual hybrid leakage may therefore propagate
through the calibration process and appear as a parallactic-angle
dependent systematic in sky-frame polarization products. This behavior
reflects the non-commuting interaction between parallactic rotation
and imperfect basis conversion.

\subsection{Regimes of Hybrid--Feed Leakage Dominance}

To understand the combined effect of $D_{\rm hyb}$ and
$D_{\rm static}$ in Equation~(\ref{eq:vrl_antenna_frame}),
three limiting regimes follow naturally:

\begin{enumerate}

\item \textbf{Hybrid-dominated leakage}
\[
|D_{\rm hyb}| \gg |D_{\rm static}|.
\]
The instrumental polarization is dominated by imperfections in
the basis-conversion network. Residual leakage is expected to be
poorly described by conventional static D-term models.

\item \textbf{Comparable leakage contributions}
\[
|D_{\rm hyb}| \sim |D_{\rm static}|.
\]
Both receptor cross-coupling and hybrid imbalance contribute
significantly to the observed leakage. Their combined effect depends
on both amplitude and relative phase and may complicate calibration of
the residual polarization response.

\item \textbf{Feed-dominated leakage}
\[
|D_{\rm hyb}| \ll |D_{\rm static}|.
\]
The leakage behaves as a conventional static D-term and standard
polarization calibration remains valid.

\end{enumerate}

The key distinction between the two leakage mechanisms is therefore
their location within the signal chain rather than their
appearance in the raw antenna-frame visibility: feed leakage
originates from cross-coupling between the linear receptors, whereas
hybrid leakage arises from imperfections in the subsequent
basis-conversion operation.

\subsection{Calibration of Hybrid Non-Idealities ($\epsilon$, $\delta$)}

Implementation of the SOP framework requires estimates of the hybrid
imperfections $\epsilon(\nu)$ and $\delta(\nu)$, which describe
frequency-dependent amplitude and phase imbalance in the
linear-to-circular conversion.

In practical systems, standard cross-hand delay calibration must first
be applied to remove differential electrical delays between the two
polarization signal chains. After the quadrature hybrid, the
pseudo-circular $R$ and $L$ voltages propagate through independent
analog paths, including cables, low-noise amplifiers (LNAs), filters,
and backend electronics before reaching the correlator. Small
differences in path length introduce a deterministic frequency-dependent
phase slope of the form
\begin{equation}
e^{-2\pi i \nu \Delta\tau},
\end{equation}
where $\Delta\tau$ is the differential delay between the two
polarization channels.

This delay is typically measured using an injected broadband
calibration signal or noise diode placed ahead of the LNAs and is
removed through standard RL/LR phase calibration. The analysis
presented here assumes that such differential-delay calibration has
already been applied, so that the remaining response is dominated by
the intrinsic hybrid imperfections
$\epsilon(\nu)$ and $\delta(\nu)$.

Two complementary approaches may be used.

\subsubsection{(i) Engineering calibration}

In the engineering approach, the hybrid is characterized in the
laboratory as a two-port RF network using vector network analyzer
(VNA) measurements. The effective response is written as
\begin{equation}
H_Y(\nu)
=
i\left[1+\epsilon(\nu)\right]
-
\delta(\nu),
\end{equation}
where deviations from ideal quadrature determine the amplitude and
phase imbalance parameters.

This provides an antenna-based prior on the instrumental response,
although it may not capture in-situ effects such as thermal drifts or
cable reflections.

\subsubsection{(ii) Astronomical self-calibration}

Alternatively, $\epsilon(\nu)$ and $\delta(\nu)$ may be estimated
directly from interferometric visibilities using polarized or
unpolarized calibrators.

For an unpolarized source, the cross-hand visibility vanishes,
\begin{equation}
V_{RL}^{\rm true}=0,
\end{equation}
so any measured cross-hand signal is instrumental:
\begin{equation}
V_{RL}^{\rm cal}
\approx
\frac{I}{2}D_{\rm eff}(\nu,\chi).
\end{equation}

Observations over a range of parallactic angles allow separation of
the static and rotating leakage contributions, enabling estimation of
$\epsilon(\nu)$ and $\delta(\nu)$ from the measured cross-hand
response.

\subsubsection{Degeneracy and calibration limitation}

A partial degeneracy exists between feed leakage and hybrid leakage,
since both contribute additive complex terms proportional to Stokes
$I$. However, the two effects differ in their parallactic-angle
behavior:

\begin{itemize}
\item Feed leakage remains static in the sky frame.
\item Hybrid leakage rotates as $e^{2i\chi}$.
\end{itemize}

Sufficient parallactic-angle coverage therefore allows the two
contributions to be separated observationally.

\subsubsection{Practical implication for SOP calibration}

The SOP correction requires only first-order accuracy in
$\epsilon(\nu)$ and $\delta(\nu)$. Residual calibration errors
propagate linearly into the corrected visibility:
\begin{equation}
\Delta V_{RL}
\sim
\frac{I}{2}
\left(
\Delta\epsilon+i\Delta\delta
\right).
\end{equation}

These residuals do not reintroduce the non-commutative coupling, but
instead determine the remaining polarization leakage floor after SOP
correction.

\subsection{Scope and Applicability of the Analysis}

The formalism developed here applies generally to systems in which 
circular polarization states are synthesized from orthogonal linear 
receptor signals downstream of parallactic rotation.

Whenever the linear-to-circular conversion operator deviates from 
the ideal unitary transformation, it does not, in general, commute 
with parallactic rotation in the linear receptor basis, producing 
the rotating leakage term derived earlier. The effect is therefore 
not specific to a single telescope or hardware implementation, but 
arises from the non-commutative interaction between parallactic 
rotation and an imperfect polarization-synthesis operator.

The analysis applies to a broad class of systems, including:

\begin{itemize}
	\item analog quadrature hybrids following orthogonal linear feeds,

	\item digital polarization synthesis applied to linear receptor voltages,

	\item OMT-based systems with downstream (analog or digital) circularization stages.
\end{itemize}

The magnitude of the effect depends on the stability, bandwidth, and
unitary accuracy of the polarization-conversion stage. Systems
employing native circular feeds largely avoid this specific geometric
coupling because the circular basis is defined directly at the feed
aperture, rather than synthesized from rotated linear receptor
signals.

The standing-wave structures discussed in Appendix C represent one
physical mechanism contributing to frequency-dependent $\epsilon(\nu)$
and $\delta(\nu)$, but the formalism applies generally to any process
introducing non-ideal linear-to-circular conversion.

The present analysis should therefore be viewed as a general framework
for polarization errors arising from non-commuting parallactic
rotation and imperfect polarization synthesis, with the GMRT hybrid
architecture serving as one concrete instrumental realization.

\section{Phased-Array Limitations with Pseudo-Circular Feeds}

The impact of pseudo-circular feed imperfections is significantly
amplified in phased-array (tied-array) systems. In this architecture,
voltages from $N$ antennas are adjusted by complex phasing weights
$w_i$ and combined into a single synthesized beam via coherent
summation. During this process, the individual antenna-frame defects
$(\epsilon_i, \delta_i)$ undergo a linear superposition, resulting in
a composite instrumental response.

The resulting measured cross-hand visibility of the phased array is:
\begin{equation}
V_{RL, \rm array}^{meas}(\nu) \approx \frac{1}{2}(Q+iU)e^{-2i\chi} - \frac{I}{2} \mathcal{D}_{\rm array}(\nu, \chi)
\end{equation}
where $\mathcal{D}_{\rm array}$ represents the weighted average of the
individual antenna leakages, where the averaging is performed in the 
complex voltage domain prior to correlation, and therefore depends on 
the phasing weights of individual antenna.  This introduces two critical problems:

\subsection{Coherent Error Accumulation}
Unlike stochastic thermal noise, which integrates down as
$1/\sqrt{N}$, quadrature hybrid defects are often systematic. If the
antennas utilize hardware from the same manufacturing batch or share
common cable-length designs, the errors $(\epsilon_i, \delta_i)$ may
be highly correlated. In this regime, the composite leakage
$\mathcal{D}_{\rm array}$ does not diminish with the addition of more
antennas; instead, it persists as a systematic floor, limiting the
polarimetric dynamic range regardless of the total collecting area.

\subsection{Non-Stationary Leakage Dynamics}
As the array tracks a source, the phasing weights $w_i$ are updated
dynamically to compensate for geometric delays. Because these weights
modulate the contribution of each antenna's unique leakage vector, the
composite leakage $\mathcal{D}_{\rm array}(\nu, \psi)$ becomes a
effectively non-stationary due to time-dependent phasing weigths.

Once the signals are summed into a tied-array beam, the individual
antenna-level leakages are no longer mathematically separable. If a
standard calibration is performed using a single leakage estimate from
a calibrator at angle $\psi_0$, the residual error becomes:
\begin{equation}
    \Delta P_{\rm residual} \approx I(\nu) \left[ \mathcal{D}_{\rm array}(\psi, \nu) - \mathcal{D}_{\rm array}(\psi_0, \nu) \right]
\end{equation}
This residual manifests as frequency-dependent EVPA wiggles that mimic
Faraday rotation. For high-fidelity results, the Static Offset
Pre-correction (SOP) must be applied to each antenna's voltage stream
prior to the beamforming summation.

\section{Discussion and Summary}

Pseudo-circular feeds were introduced as a practical solution for
low-frequency radio interferometry, motivated by the desire to combine
the mechanical simplicity of linear receptors with the calibration
advantages of circular polarization. By converting linear voltages
into approximate circular states, these systems reduce the direct
mixing of Stokes $Q$ and $U$ caused by ionospheric Faraday rotation
and allow polarization calibration to be performed in a framework
similar to that of native circular feeds. Historically, this approach
has been widely adopted across major facilities: the Very Large Array
(VLA) employs a circular basis for frequencies above 1 GHz to reduce
the impact of parallactic angle rotation \citep{Napier1983,
Perley2011}, while the Giant Metrewave Radio Telescope (GMRT)
implemented analog quadrature hybrids in its legacy meter-wavelength
bands (150--610 MHz) for pulsar and ionospheric studies
\citep{Swarup1991}. The uGMRT retains this architecture for broader
instantaneous bandwidths, whereas arrays like the Australia Telescope
Compact Array (ATCA) use native linear feeds, avoiding hybrid-induced
imperfections but requiring more complex, time-dependent leakage
modeling \citep{Sault1991, Wilson2011}.

The analysis presented here shows that while pseudo-circular feeds
offer practical calibration advantages, this approximation breaks down
in the presence of realistic hybrid imperfections. The essential issue
is that the quadrature hybrid operates in the antenna frame after the
sky polarization has been rotated by parallactic angle. As a result,
the hybrid transformation does not commute with geometric rotation. A
static hardware imperfection therefore appears in the sky frame as a
time-dependent leakage term, introducing a systematic coupling between
total intensity and linear polarization that cannot be fully removed
by standard polarization calibration procedures (SCP).

The magnitude of the resulting error depends strongly on fractional
polarization. For a source with fractional polarization $m$, the
intrinsic polarized signal scales as $mI$, while the leakage scales as
$|D|I$. The induced EVPA error therefore scales approximately as
$|D|/m$, implying that weakly polarized sources are particularly
susceptible. In this regime the instrumental contribution can
dominate the observed polarization, producing non-linear EVPA
structure and biased rotation-measure estimates.

Simulations presented in this work confirm these expectations. When
calibration is applied after parallactic-angle correction, residual
hybrid leakage produces systematic distortions in the EVPA spectrum
and incorrect rotation measures. In contrast, removing the hybrid
contribution in the antenna frame prior to derotation --- the
Static-Offset Pre-correction (SOP) approach --- restores the expected
$\lambda^2$ behavior and substantially suppresses the leakage-induced
bias. This demonstrates that the dominant limitation arises from the
ordering of transformations in the signal chain rather than from the
absolute leakage amplitude alone.

In practice, however, implementing SOP requires accurate knowledge of
the hybrid response matrix $\mathbf{H}(\nu)$, and determining this
response in situ is non-trivial. Once embedded within a real
instrumental system, the hybrid leakage becomes coupled to intrinsic
feed leakage, standing-wave structure, bandpass variations, cable
reflections, and other antenna-dependent effects. The effective
instrumental leakage observed in the visibilities therefore represents
a superposition of multiple non-idealities rather than an isolated
hybrid contribution. As a result, separating hybrid-induced leakage
from conventional feed D-terms may itself require wide parallactic
angle coverage, stable calibration sources, or independent engineering
measurements of the hybrid response.

This highlights an important practical trade-off. Reverting to a pure
linear-feed basis removes the non-commutative hybrid conversion stage
and simplifies the instrumental polarization formalism, since all
leakage terms remain naturally defined in the antenna-frame linear
basis. However, at low radio frequencies this reintroduces the strong
time- and frequency-dependent mixing of Stokes $Q$ and $U$ caused by
ionospheric Faraday rotation, eliminating one of the primary
motivations for pseudo-circular architectures. Future wide-band
low-frequency systems may therefore require alternative calibration
strategies that preserve the advantages of circularized feeds while
avoiding the non-commutative leakage pathology identified here.

From an instrumentation perspective, reducing hybrid imperfections
provides only a partial solution. Achieving sub-percent amplitude and
phase balance over wide fractional bandwidths at low frequencies is
technically difficult due to dispersion, fabrication tolerances, and
frequency-dependent reflections. The problem is further amplified in
phased-array systems, where antenna-dependent leakage terms combine
during coherent beamforming. Once voltages are summed, individual
antenna leakages are no longer independently recoverable, imposing a
fundamental limit on polarization fidelity unless corrections are
applied prior to phasing.

These results highlight a fundamental trade-off in the use of
pseudo-circular feeds. While historically and operationally
advantageous, they introduce a calibration systematic that is
intrinsically linked to the non-commutative structure of the signal
chain. Accurate recovery of broadband polarization therefore requires
calibration strategies that explicitly account for this effect, or
alternatively, receiver architectures that avoid imperfect
linear-to-circular conversion altogether.

\begin{acknowledgments}
	We thank the anonymous referee for their valuable comments
	and constructiv suggestions that improved the paper
	significantly.
We thank colleagues from NCRA and GMRT for extensive technical
discussion. We thank the staff of the GMRT that made observations 
that helped understanding the problem presented in this paper. 
GMRT is run by the National Centre for Radio Astrophysics of the 
Tata Institute of Fundamental Research.  We thank Dharam Vir Lal 
and Subhashis Roy for careful reading of the manuscript and useful
comments. D.M. acknowledges the support of the Department of 
Atomic Energy, Government of India, under project No. 
12-R\&D-TFR-5.02-0700.
\end{acknowledgments}

\appendix

\section{Visibility Expansions in the Pseudo-Circular Basis}
\label{app_c}

The linear feed voltages at antenna $A$ are
\begin{equation}
X^A = G_{XA}(E_X + D_{XA}E_Y), \qquad
Y^A = G_{YA}(E_Y + D_{YA}E_X),
\end{equation}
with $|D_{XA}|, |D_{YA}| \ll 1$.

The quadrature hybrid converts these into pseudo-circular voltages:
\begin{equation}
\begin{pmatrix}
R^A \\
L^A
\end{pmatrix}
=
\frac{1}{\sqrt{2}}
\begin{pmatrix}
1 & H_Y \\
1 & -H_Y
\end{pmatrix}
\begin{pmatrix}
X^A \\
Y^A
\end{pmatrix},
\end{equation}
where
\begin{equation}
H_Y \approx i(1+\epsilon_A) - \delta_A.
\end{equation}

Substituting the feed model yields
\begin{equation}
R^A = \frac{1}{\sqrt{2}} \left[
G_{XA}(E_X + D_{XA}E_Y)
+ H_Y G_{YA}(E_Y + D_{YA}E_X)
\right],
\end{equation}

Grouping terms in $(E_X, E_Y)$ gives the compact form
\begin{equation}
R^A = \frac{1}{\sqrt{2}} \left(\alpha^A E_X + \beta^A E_Y \right),
\end{equation}
with
\begin{align}
\alpha^A &= G_{XA} + H_Y G_{YA} D_{YA},\\
\beta^A  &= G_{XA} D_{XA} + H_Y G_{YA}.
\end{align}

The corresponding coefficients for $L^A$ are denoted
$\tilde{\alpha}^A$ and $\tilde{\beta}^A$, and are obtained from
$\alpha^A$ and $\beta^A$ under the substitution
\begin{equation}
H_Y \rightarrow -H_Y.
\end{equation}

Substituting these definitions into the measurement equation, and 
using the antenna-frame coherency terms representing the correlations 
at the native linear receptors prior to hybrid conversion:
\begin{align}
X_{11}
&=
\frac12
\left(
I + Q\cos2\chi + U\sin2\chi
\right), \\
X_{22}
&=
\frac12
\left(
I - Q\cos2\chi - U\sin2\chi
\right), \\
X_{12}
&=
\frac12
\left(
U\cos2\chi - Q\sin2\chi + iV
\right).
\end{align}
where $X_{21}=X_{12}^{\ast}$, and
the full pseudo-circular visibilities including hybrid imperfections and feed leakage are:
\begin{align}
V_{RR} &= \alpha^A \alpha_B^\ast X_{11} + \alpha^A \beta_B^\ast X_{12} + \beta^A \alpha_B^\ast X_{21} + \beta^A \beta_B^\ast X_{22},\\
V_{RL} &= \alpha^A \tilde{\alpha}_B^\ast X_{11} + \alpha^A \tilde{\beta}_B^\ast X_{12} + \beta^A \tilde{\alpha}_B^\ast X_{21} + \beta^A \tilde{\beta}_B^\ast X_{22},\\
V_{LR} &= \tilde{\alpha}^A \alpha_B^\ast X_{11} + \tilde{\alpha}^A \beta_B^\ast X_{12} + \tilde{\beta}^A \alpha_B^\ast X_{21} + \tilde{\beta}^A \beta_B^\ast X_{22},\\
V_{LL} &= \tilde{\alpha}^A \tilde{\alpha}_B^\ast X_{11} + \tilde{\alpha}^A \tilde{\beta}_B^\ast X_{12} + \tilde{\beta}^A \tilde{\alpha}_B^\ast X_{21} + \tilde{\beta}^A \tilde{\beta}_B^\ast X_{22}.
\end{align}

\section{Physical Motivation of the Hybrid Error Model}
\label{app_d}

To accurately simulate the impact of quadrature hybrid imperfections
on polarimetric recovery, we adopt a functional form for the amplitude
imbalance $\epsilon(\nu)$ and phase error $\delta(\nu)$ that is
grounded in the electromagnetic behavior of distributed-element
circuits. In the frequency range of 500--1000~MHz, such as the Band4 in 
uGMRT, these components
are typically implemented as branch-line couplers, whose response is
governed by the interference of signals propagating through
transmission line segments of finite electrical length.

\subsection{The Ideal Quadrature Response}

The response of a quadrature hybrid is defined by a complex transfer
function $H(\nu)$, determined by the electrical lengths of its
internal transmission line segments. For a hybrid with characteristic
impedance $Z_0$, the response depends on the electrical length
$\varphi$ of the transmission paths. The electrical length is defined
by the physical length $L$ of the hybrid arms and the observing
frequency $\nu$:
\begin{equation}
\varphi(\nu) = \frac{2\pi \nu L}{v_{p}},
\end{equation}
where $v_{p}$ is the phase velocity in the dielectric medium. An ideal
hybrid is designed such that at the center frequency $\nu_0$, the
length $L$ corresponds to a quarter-wavelength ($\lambda/4$), yielding
$\varphi(\nu_0) = \pi/2$ and an ideal quadrature phase shift of
$+90^\circ$.

\subsection{Taylor Expansion of Structural Dispersion}

In wide-band applications, the observing frequency deviates from the
design center by $\Delta \nu = \nu - \nu_0$. We define the fractional
frequency offset as $x = (\nu - \nu_0)/\nu_0$, such that the
electrical length of the hybrid arms scales as $\varphi(x) =
\frac{\pi}{2}(1 + x)$. To model the smooth baseline deviation from
ideal behavior, we expand the complex hybrid error response $\Delta
H(x)$ about the design point $x = 0$:
\begin{equation}
\Delta H(x) \approx \Delta H(0) + \Delta H'(0)x + \frac{1}{2}\Delta H''(0)x^2 + \frac{1}{6}\Delta H'''(0)x^3 + \mathcal{O}(x^4).
\end{equation}

For a ideal branch-line four-port quadrature hybrid, 
symmetry implies that the error response $\Delta H(x)$ is approximately 
odd with respect to frequency offset, $\Delta H(x) = -\Delta H(-x)$
(\citealt{pozar2011microwave}), so that even-order terms are strongly
suppressed. In practice, fabrication tolerances, finite isolation, and
integration into a full receiver chain break this symmetry, producing
a non-zero static term $\Delta H(0)$ and allowing residual even-order
contributions. In our phenomenological model, these residual even-order
terms are assumed to be slowly varying and are absorbed into standard
bandpass calibration, leaving the lowest-order non-trivial dispersive
structure as the dominant uncalibrated effect.

Furthermore, well-designed quadrature hybrids are engineered for
maximal flatness at the design frequency, implying that
$\Delta H'(0) \approx 0$ over the usable fractional bandwidth.
Consequently, the leading frequency-dependent term is cubic, and we
parameterize the hybrid dispersion as

\begin{equation}
\Delta H_{\rm base}(\nu)
= \Delta H_0
+ A_\epsilon \left(\frac{\nu - \nu_0}{\nu_0}\right)^3
+ i\,A_\delta \left(\frac{\nu - \nu_0}{\nu_0}\right)^3.
\end{equation}

Here $\Delta H_0$ is the static complex imbalance arising from
manufacturing tolerances, while $A_\epsilon$ and $A_\delta$ are
dimensionless coefficients describing the leading-order cubic
frequency dependence of the in-phase and quadrature error components,
respectively. This form corresponds directly to the implementation in
the simulation code, where the hybrid model is constructed as a smooth
cubic baseline plus additional sinusoidal ripple terms.

\subsection{Standing Waves from Impedance Mismatch}

In addition to smooth structural dispersion, real receiver chains
exhibit frequency-dependent ripple due to impedance mismatches between
the hybrid, cabling, and low-noise amplifiers. These effects are
well-described by reflection coefficients $\Gamma_k$ at discrete
electrical path delays $\tau_k$.

A reflected signal produces a coherent superposition
\begin{equation}
E_{\rm meas}(\nu)
=
E_0 \left[ 1 + \Gamma e^{-i(2\pi \nu \tau + \phi)} \right],
\end{equation}
leading to a sinusoidal modulation in frequency space. For multiple
reflection paths, the resulting instrumental contribution is

\begin{align}
\epsilon_{\rm wig}(\nu) &= \sum_k M_k \cos(2\pi \nu \tau_k + \phi_k), \\
\delta_{\rm wig}(\nu) &= \sum_k P_k \sin(2\pi \nu \tau_k + \psi_k).
\end{align}

These terms correspond directly to the sinusoidal “wiggles” implemented
in the simulation code and represent non-polynomial, non-smooth
instrumental structure that cannot be captured by Taylor expansion.

\subsection{Unified Functional Form for Hardware Emulation}

The full complex hybrid error is written as
\begin{equation}
\Delta H(\nu) = \epsilon(\nu) + i\,\delta(\nu),
\end{equation}
with
\begin{align}
\epsilon(\nu) &= \Delta H_{\epsilon,0}
+ A_\epsilon x^3
+ \epsilon_{\rm wig}(\nu), \\
\delta(\nu) &= \Delta H_{\delta,0}
+ A_\delta x^3
+ \delta_{\rm wig}(\nu),
\end{align}
where $x = (\nu - \nu_0)/\nu_0$.

Thus,
\begin{equation}
\Delta H(\nu)
=
\Delta H_0
+
(A_\epsilon + i A_\delta)x^3
+
\sum_k |\Gamma_k| e^{i(2\pi \nu \tau_k + \phi_k)}.
\end{equation}

Manufacturers typically specify a static amplitude imbalance
$\Delta H_0$ of approximately 0.5~dB for standalone quadrature hybrids.
However, the integration of these components into a complete receiver
chain—involving coaxial interconnects, low-noise amplifiers (LNAs), and
atmospheric protection housings—introduces additional impedance
mismatches and gain/phase asymmetries. These system-level effects
can readily increase the effective static imbalance to the
0.8--1.0~dB range (approximately $\sim 10\%$ in voltage).

In our simulations, we therefore adopt a conservative baseline offset
$\Delta H_0 = 0.1$, ensuring that the Static Offset Pre-correction (SOP)
method is evaluated under a realistic worst-case instrumental
leakage regime rather than an idealized laboratory specification.

\subsection{Connection to the Numerical Hybrid Model}

The simulation implements this model as
\begin{align}
\epsilon(\nu) &= \epsilon_0 + A_\epsilon x^3 + \epsilon_{\rm wig}(\nu), \\
\delta(\nu)   &= \delta_0 + A_\delta x^3 + \delta_{\rm wig}(\nu),
\end{align}
where $x = (\nu - \nu_0)/\nu_0$, and the oscillatory terms
$\epsilon_{\rm wig}$ and $\delta_{\rm wig}$ represent standing-wave
modulations due to impedance mismatches in the receiver chain.

In this framework, the complex hybrid defect is
\begin{equation}
\Delta H(\nu) = \epsilon(\nu) + i\,\delta(\nu).
\end{equation}

\subsection{Consistency with Static Offset Pre-correction (SOP)}

The measured visibility in the antenna frame (see Eq.~\ref{eq:sop_meas}) is written as
\begin{equation}
V_{RL}^{\rm meas}(\nu,\chi)
= \frac{1}{2}V_{RL}^{\rm true}(\nu)e^{-2i\chi}
	- \frac{I}{2}\,D_{\rm hyb}(\nu).
\end{equation}

where $\chi$ is the parallactic angle and $I$ is the total intensity.
Here $D_{\rm hyb}(\nu)=\Delta H(\nu)$ represents the full complex
antenna-frame hybrid response.
The SOP procedure explicitly removes the full antenna-frame hybrid
response:
\begin{equation}
V_{RL}^{\rm SOP}(\nu)
= V_{RL}^{\rm meas}(\nu,\chi)
+ \frac{I}{2}\,\Delta H(\nu).
\end{equation}

After correction, the sky term is restored exactly in this model, while
the key failure mode of standard calibration arises when only a
frequency-independent or parallactic-angle–averaged estimate of
$\Delta H$ is removed. The residual error is therefore governed by the
frequency-dependent mismatch
$\Delta H(\nu) - \Delta H(\nu_0)$, which is dominated by the cubic
dispersion and ripple terms described above.

This makes the role of SOP explicit: it is not merely a gain
correction, but a removal of the full voltage-level complex defect
$\Delta H(\nu)$ prior to geometric rotation.

\subsection{Physical Interpretation and Method Justification}
The interpretation follows directly from the visibility equation in the antenna frame,
where $\Delta H(\nu)$ enters as an additive, non-rotating complex leakage term
that is subsequently mixed with the parallactic angle rotation $e^{-2i\chi}$.
This functional form of $\Delta H(\nu)$ used above is significant for two key reasons:

\begin{enumerate}
    \item Geometric Distortion: The complex error $\Delta H(\nu)$ acts
      as a static vector in the antenna frame. When the sky signal is
      de-rotated by $e^{2i\chi}$, the relative orientation between the
      sky's polarization vector and the fixed hardware defect
      changes. If $\epsilon \neq \delta$, the leakage traces an
      elliptical path in the $(Q,U)$ plane, creating a dynamic bias
      that standard calibration---which assumes a circular or constant
      response---cannot mitigate.
    \item Temporal Stability: While reflection coefficients
      $|\Gamma_k|$ may exhibit minor thermal fluctuations, the
      structural dispersion $A$ and the fundamental standing-wave
      frequencies $\tau_k$ are determined by the fixed physical
      geometry of the etched circuits and cable lengths. This
      stability justifies the SOP method's reliance on a one-time
      high-fidelity calibration of the hybrid response to solve for
      the hardware ``signature'' across the band.
\end{enumerate}

Critically, because these errors are defined at the voltage level in
the antenna frame, the Static Offset Pre-correction (SOP) method
operates on the full $\Delta H(\nu)$ term prior to parallactic-angle
rotation and beamforming. This removes both the static and
frequency-dependent components of the hybrid response in their native
frame, preventing them from mixing with the geometric rotation term
$e^{2i\chi}$.

\bibliography{refs}{}

\bibliographystyle{aasjournal}

\end{document}